\documentclass[aps,prl,twocolumn,floatfix,superscriptaddress]{revtex4}  
\usepackage{float}
\usepackage{graphicx}
\usepackage{amssymb}
\usepackage{dsfont}
\usepackage{amsmath}  
\usepackage{epstopdf}
\usepackage{xcolor}
\begin{document}

\def\ket#1{|#1\rangle} 
\def\bra#1{\langle#1|}
\def\av#1{\langle#1\rangle}
\def\dkp#1{\kappa+i(\Delta+#1)}
\def\dkm#1{\kappa-i(\Delta+#1)}
\def\pp{{\prime\prime}}
\def\ppp{{\prime\prime\prime}}
\def\w{\omega}
\def\k{\kappa}
\def\D{\Delta}
\def\wp{\omega^\prime}
\def\wpp{\omega^{\prime\prime}}

\title{Large collective power enhancement in dissipative charging of a quantum battery}
\author{Sagar Pokhrel and Julio Gea-Banacloche}
\affiliation{University of Arkansas, Fayetteville, AR}

\date{\today}

\begin{abstract}
We consider a model for a quantum battery consisting of a collection of $N$ two-level atoms driven by a classical field and decaying to a common reservoir.  In the extensive regime, where the energy $E$ scales as $N$ and the fluctuations $\Delta E/E \to 0$, our dissipative charging protocol yields a power proportional to $N^2$, a scaling that cannot be achieved in this regime by Hamiltonian protocols.  The tradeoff for this enhanced charging power is a relative inefficiency, since a large fraction of the incoming energy is lost through spontaneous emission to the environment.  Nevertheless,  we find the system can store a large amount of coherence, and also release the stored energy coherently through spontaneous emission, again with a power scaling as $N^2$. 
\end{abstract}
\maketitle

{\it Introduction.---}``Quantum batteries'' \cite{campaioli} are energy-storage systems described by quantum mechanics that may exploit quantum resources, such as coherence or entanglement, to achieve an advantage over classical batteries: for example, collective manipulation of a set of quantum batteries may increase their extractable energy (or ``ergotropy'') \cite{alicki}, or their charging (or discharging) power \cite{binder,campaioli2}.  As with other quantum thermodynamics tasks  \cite{kurizki, petruccione, fazio, schaller,tokura,kim}, much attention has been paid to systems that rely on superradiance \cite{dicke,haroche}, or, more generally, on a collective enhancement arising from the totally symmetric manipulation and evolution of a collection of two-level systems (or qubits).  For these so-called ``Dicke batteries'' \cite{ferraro,seidov,quach}, in the absence of direct interaction between the qubits and in the extensive regime, it has been shown \cite{campaioli2,julia} that for Hamiltonian charging schemes the collective enhancement is limited to a factor $\sqrt N$, giving a total power that scales as $N^{3/2}$.  Here, we show how a dissipative charging protocol can, in principle, achieve an enhancement factor of $N$, that is, a total power that scales as $N^2$, in the same regime. (See \cite{barra,roncaglia,zakavati} for other recent studies of the potential advantages of dissipative charging schemes.)

{\it Charging scheme.---}  Our system is a collection of $N$ two-level atoms, and our  charging procedure is a direct extension of the one-atom state preparation scheme described in \cite{monsel,monsel2}.  The atoms should be well within a wavelength of each other; they are ``dressed'' (and ultimately get their energy from) a strong ``pump'' field that we treat classically, and they decay, collectively, to the same environment.  Importantly, this could be carried out, in principle, even in free space, and still exhibit the same collective advantage; nevertheless, in order to take full advantage of the method's potential (a point to which we will return later) some degree of environment manipulation is desirable, which could be achieved by confining the atoms in waveguides, optical cavities, or photonic crystals.  

By choosing the zero of energy appropriately, and introducing collective angular momentum operators to describe the collection of $N$ two-level systems (each of which can be considered as an individual ``battery''), the system alone can be described by a Hamiltonian $H_0 = \omega_0 J_z$.  In the presence of the external field, we  write instead
\begin{equation}
H_1=\Delta J_z -\frac{\Omega_R}{2}\left(J_++J_-\right) = \Delta J_z -{\Omega_R}J_x
\label{n1}
\end{equation}
where $\Omega_R$ is the driving field's Rabi frequency and $\Delta = \omega_0 - \omega_P$ its detuning ($\omega_0$ is the atomic frequency and $\omega_P$ the pump frequency).  In terms of the individual atoms' excited and ground states, $\ket e$ and $\ket g$, the operators $J_z$ and $J_+$ are 
\begin{equation}
J_z = \frac \hbar 2 \sum_{i=1}^N \left(\ket e_i\bra e - \ket g_i \bra g \right), \quad J_+ = \hbar \sum_{i=1}^N \ket e_i\bra g.
\label{e2}
\end{equation}
 Defining $\Omega_P = \sqrt{\Delta^2 + \Omega_R^2}$ and introducing the angle $\theta = \tan^{-1}(\Omega_R/\Delta)$, with $0\le\theta\le\pi$, we can rewrite $H_1$ as
\begin{equation}
H_1 = \Omega_P  J_z^\prime
\label{n3}
\end{equation}
where $J_z^\prime = \cos\theta J_z-\sin\theta  J_x = e^{i J_y\theta/\hbar}  J_z e^{-i J_y\theta/\hbar} $ is just $J_z$ rotated  by an angle $\theta$ clockwise around the $y$ axis. The eigenstates of $H_1$, which we will denote by $\{\ket{e_m}\}$ (the ``dressed state'' basis), will therefore  be the rotated eigenstates $\{\ket m\}$ of $J_z$:
\begin{equation}
\ket{e_m} = e^{i J_y\theta/\hbar}\ket m
\label{n5}
\end{equation}
with the corresponding eigenenergies $ m \hbar \Omega_P$, $-N/2 \le m \le N/2$.  We will use primes to denote any rotated operator in what follows.

To include decay through spontaneous emission, we take as our starting point the standard master equation, in Lindblad form, which, assuming all the atoms decay to a common reservoir, would be written \cite{carmichael}
\begin{align}
\dot\rho &=-\frac{i}{\hbar}[H_1,\rho] + \frac{\gamma}{\hbar^2} {\cal L}[J_-]\rho \cr
&= -\frac{i}{\hbar}[H_1,\rho] +\frac{\gamma}{\hbar^2} \left [J_- \rho J_+ - \frac 1 2 \left( J_+ J_-\rho +\rho J_+ J_- \right) \right]
\label{n8}
\end{align}
We note, however, that the collective decay operator $J_-$ has the decomposition, in the rotated basis
\begin{equation}
J_- = -\sin\theta J^\prime_z -\frac{1-\cos\theta}{2} J^\prime_+ + \frac{1+\cos\theta}{2} J^\prime_- 
\end{equation}
and that, under the action of the Hamiltonian (\ref{n3}), the collective dipole operators $J^\prime_+$ and $J^\prime_-$ (proportional to sums of terms of the form $\ket{e_{m+1}}\bra{e_m}$ and $\ket{e_{m-1}}\bra{e_m}$, respectively) will evolve as $e^{i\Omega_P t}$ and $e^{-i\Omega_P t}$, respectively.  If we keep only secular terms in the master equation, and include the evolution due to $H_1$, we find that (\ref{n8}) breaks up into three pieces, each one proportional to (in principle) a different decay rate $\gamma$, which depends on the density of modes at the corresponding frequency:
\begin{align}
\frac{d}{dt}\rho = &-\frac{i}{\hbar}\left[H_1,\rho\right] + \frac{\gamma_0}{\hbar^2}\sin^2\theta {\cal L}[J_z^\prime]\rho \cr
&+ \frac{\gamma_-}{\hbar^2}\sin^4\left(\frac \theta 2\right){\cal L}[J_+^\prime]\rho + \frac{\gamma_+}{\hbar^2}\cos^4\left(\frac \theta 2\right){\cal L}[J_-^\prime]\rho \cr
\label{n10}
\end{align}
where $\gamma_\pm \equiv \gamma(\omega_P \pm \Omega_P)$, and $\gamma_0 \equiv \gamma(\omega_P)$.
The decay terms in (\ref{n10}) correspond to spontaneous emission of photons at the frequencies $\omega_p$ and $\omega_p \pm \Omega_p$ of a Mollow triplet \cite{mollow}.  Their relative rates can be modified either by changing the detuning, or ``engineering the environment,'' i.e., modifying the $\gamma_\pm$ rates \footnote{See \cite{monsel2} for further details; note, however, that their $\Sigma_+$ operators correspond to our $J_-$, and likewise $\Sigma_-$ to $J_+$}. The idea behind the charging process described by Eq.~(\ref{n10}) is, essentially, that spontaneous emission in the dressed basis can actually move the system up the energy ladder in the ``bare'' basis $\{\ket m\}$, and this process is enhanced by the collective factor $N$.

The steady state of (\ref{n10}) is a diagonal matrix in the $\{\ket{e_m}\}$ basis, whose elements can be found using the detailed balance condition 
$\gamma_-\sin^4(\theta/2) \rho_{m,m} = \gamma_+\cos^4(\theta/2) \rho_{m+1,m+1}$.  They can be written as
\begin{equation}
\rho^{(ss)}_{m,m} = \frac{x^{N/2-m}}{\sum_{n=0}^N x^n}
\label{n13}
\end{equation}
where we have defined $
x = r \cot^4\left(\frac \theta 2 \right)$ and 
$r = \gamma_+/\gamma_-$.
From this, in turn, making use of the rotation operation (\ref{n5}), we obtain the system's energy
\begin{equation}
Tr\left(\rho^{(ss)} H_0\right) =  \frac{\hbar\omega_0\cos\theta}{\sum_{n=0}^N x^n} \sum_{m=-N/2}^{N/2} m x^{N/2-m}
\label{e11}
\end{equation}
and its ergotropy $R= Tr\left(\rho^{(ss)} H_0\right) - Tr\left(\rho_p H_0\right)$, where the \emph{passive state} density matrix $\rho_p$ is constructed by assigning to the eigenvalues of $H_0$, in decreasing order, the probabilities (\ref{n13}), in increasing order.  We therefore have to distinguish the two cases $x>1$ and $x<1$, with the result,
 for the ergotropy per atom (after evaluating the sums in Eq.~(\ref{e11}) in closed form),
\begin{equation}
\frac R N = \frac{\hbar\omega_0}{2N}(\cos\theta\pm 1)\frac{N(x-1)(1+x^{N+1})+2 x(1-x^N)}{(x-1)(1-x^{N+1})}
\label{e24}
\end{equation}
where the $+$ sign applies for $x<1$, and the $-$ for $x>1$.
This is plotted in Figure 1, as a function of $\theta$, for $r\equiv \gamma_+/\gamma_- = 0.1$, and for two different values of the number of atoms $N$.  Note that, for these small values of $N$, the ergotropy scales superextensively. 
\begin{figure}
    \includegraphics[width=8cm]{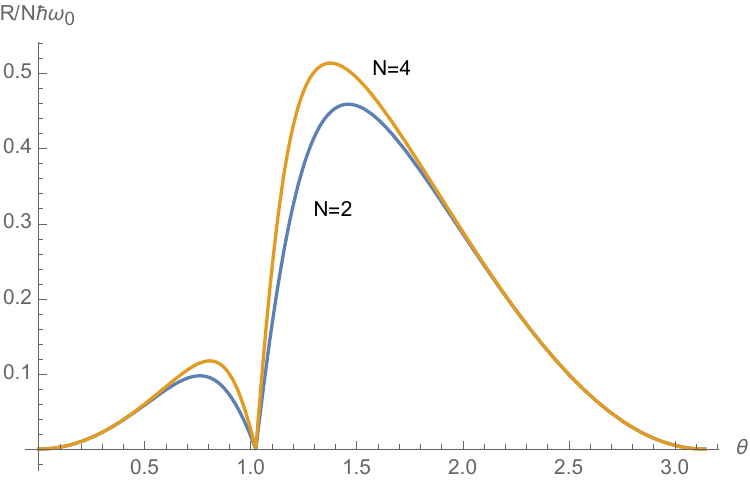}%
    {\caption{ Ergotropy per atom, as a function of $\theta$, for $N=2$ and $N=4$, and $r=0.1$. Note that the ergotropy always goes to zero for $x=1$, since in that case all the levels have the same occupation probability, and $\rho_{ss} = \rho_p$. The left side of the figure corresponds to $x>1$ and the right side to $x<1$; for $r=1$ the figure would be symmetric, but for $r<1$, as is the case here (meaning $\gamma_+<\gamma_-$) it is clearly advantageous to operate with $x<1$, while the converse is the case for $r>1$. }}
\label{fig:fig1}
\end{figure}

{\it The extensive regime.---}  For large $N$ and $x<1$, $x^N$ in Eq.~(\ref{e24}) eventually approaches zero and we get 
\begin{equation}
\frac R N \to \hbar\omega_0 \cos^2(\theta/2) = \frac{\hbar\omega_0}{1+\sqrt{r/x}} \quad (x<1)
\label{e25}
\end{equation}
Since this expression assumes $x<1$, we see the ergotropy per atom cannot exceed $\hbar\omega_0/(1+\sqrt r)$.  Note that, in this limit, we also find $\av{H_0} + N\hbar\omega_0/2 = R$, that is, assuming the battery starts in the ground state, all the energy deposited in it by the charging process becomes extractable (in agreement with the general predictions in \cite{andolina}). 

Conversely, for $x>1$, $x^{N+2}$ in (\ref{e24}) eventually dominates and we get 
\begin{equation}
\frac R N \to \hbar\omega_0 \sin^2(\theta/2) = \frac{\hbar\omega_0}{1+\sqrt{x/r}} \quad (x>1)
\label{e26}
\end{equation}
which agains satisfies $R = \av{H_0} + N\hbar\omega_0/2$, and now is bounded by $\hbar\omega_0/(1+\sqrt{1/r})$.  For $r=1$ ($\gamma_+ = \gamma_-$), both bounds coincide, and the maximum ergotropy per atom is then $\hbar\omega_0/2$. Hence, some environment engineering (i.e., making $\gamma_+\ne\gamma_-$)  is required if we want to charge the battery to more than half capacity.  It is easy to see that, for $x<1$, one can have $R/N \to \hbar\omega_0$ in the limit $r\ll x < 1$, while for $x>1$ this happens for $1<x\ll r$.

We consider next the dynamics of the charging process, where the battery charge is given by $\av{E} =\av{H_0} + N\hbar\omega_0/2 =   \omega_0 Tr(\rho J_z) + N\hbar\omega_0/2$. With $\rho$ in the rotated basis, we have
\begin{align}
Tr(\rho J_z)  &=Tr(\rho e^{-i J_y\theta/\hbar} J_z ^\prime e^{i J_y\theta/\hbar}) \cr
&= \cos\theta \,Tr(\rho J_z^\prime)  + \sin\theta \,Tr(\rho J_x^\prime )
\label{e13}
\end{align}
The time evolution of (\ref{e13}) can then be calculated from Eq.~(\ref{n10}).  Assuming the energy uncertainty, $\Delta E$, is small compared to $\av E$, we  obtain \cite{supp} the following analytical result, valid for large $N$.
\begin{align}
\frac{1}{N\hbar \omega_0} \av{E}& = \frac 1 2 -\frac{1}{2} \cos\theta\tanh\left[\frac N 2\Gamma t + \phi_0 \right] \cr 
&-\frac 1 2 \sin\theta\frac{\cos(\Omega_P t) }{\cosh\left(N\Gamma t /2+ \phi_0\right)}\label{e15}
\end{align}
with $\phi_0 =  \tanh^{-1}(\cos\theta) = \ln|\cot(\theta/2)|$ and
\begin{equation}
\Gamma = \gamma_+\cos^4 \frac \theta 2 - \gamma_-\sin^4 \frac \theta 2 
\label{n15}
\end{equation}
(note that $\Gamma <0$ if $x<1$;  equation (\ref{e15}) still holds in this case, but for simplicity we will restrict ourselves in what follows to the $x>1$ case, and to $r\ge 1$).

We note that the secular approximation we have used to recast (\ref{n8}) in the form (\ref{n10}) requires $\Omega_P \gg \gamma_+,\gamma_-,\gamma_0$, and we have also found that, in general, it does not predict the steady state reliably as $N$ increases unless $\Omega_P$ also increases with $N$ (i.e., we need $\Omega_P \gg N\Gamma$ as well).  Physically, this means the power in the driving field must increase as $N^2$, which makes sense, since one should not expect the battery to charge faster than the source can supply the energy.  Formally, this means the oscillations in (\ref{e15}) persist in the large $N$ limit. 
\begin{figure}
    \includegraphics[width=8.5cm]{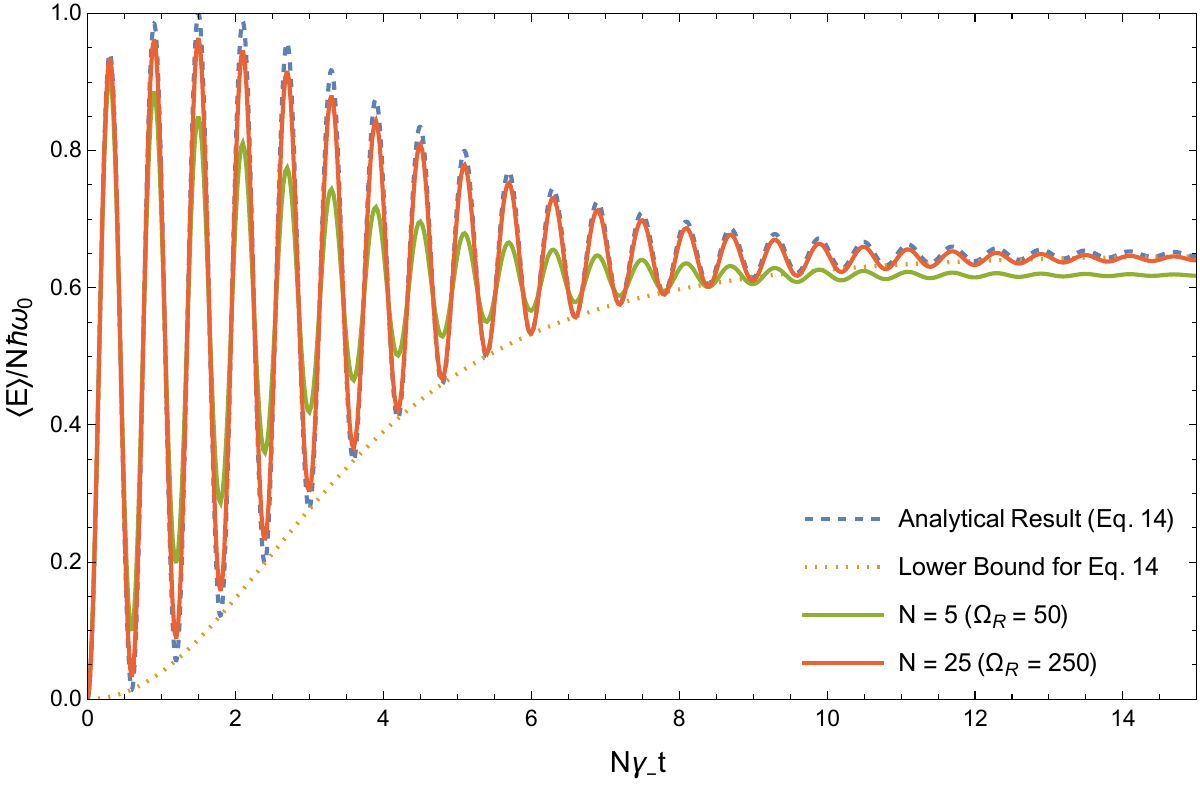}%
    {\caption{Solid lines: average energy per qubit as a function of $N\gamma_-t$, from numerical integration of Eq.~(\ref{n10}), with parameters  $r = 10$, $\theta = 1.87$, $\gamma_0=\gamma_-$, and $N$ and $\Omega_R$ as shown.  Dashed line: the asymptotic analytical solution (\ref{e15}).  Dotted line: lower bound to the analytical solution.}}
\label{fig:fig2}
\end{figure} 

It is clear from (\ref{e15}) that the instantaneous power, $P(t)= d\av{E}/dt$, scales as $N^2$, although with large oscillations before the steady state is reached.  It is customary to define an average power as $P_{av} = (1/T)\int_0^T P(t)dt = \av{E(T)}/T$ for some time interval $T$, but as this is also an oscillating quantity it is important to choose the time $T$ appropriately.  Examination of (\ref{e15}) shows that a lower bound to $\av{E(t)}$ can be obtained by replacing $\cos(\Omega_P t)$ by 1, i.e., $\av{E(t)} \ge N\hbar\omega_0(1-\cos\theta\tanh(\tau+\phi_0)-\sin\theta /\cosh(\tau+\phi_0))/2$, with $\tau = N\Gamma t/2$.  Let $\tau_{90}$ be the (dimensionless) time when this expression reaches $90\%$ of its steady-state value, $N\hbar\omega_0\sin^2(\theta/2)$, for a given $\theta$.  As our lower bound is a monotonic function of $\tau$, we know that the battery charge will never oscillate below this value for $\tau>\tau_{90}$.  Choosing then $T=2\tau_{90}/N\Gamma$ for our definition of $P_{av}$, we obtain the following lower bound:
\begin{equation}
P_{av} \ge \frac{0.9\sin^2(\theta/2)}{2\tau_{90}}N^2\hbar\omega_0\Gamma
\end{equation}
Using (\ref{n15}), this can be rewritten as 
\begin{equation}
\frac{P_{av}}{N^2\hbar\omega_0\gamma_-} \ge \frac{0.9}{2\tau_{90}}{\cal E}\left(r(1-{\cal E})^2 - {\cal E}^2 \right)
\label{e18}
\end{equation}
where ${\cal E} = \sin^2(\theta/2)$ is the final battery charge, divided by $N\hbar\omega_0$.  Figure 3 shows a sample plot of this relationship, which clearly indicates a tradeoff between average charging power and final battery charge, for a given $r>1$.  [For $x$ and $r$ less than 1, the result corresponding  to (\ref{e18}) has $ {\cal E}\left((1-{\cal E})^2 - r{\cal E}^2 \right)$ on the right-hand side, with ${\cal E} = \cos^2(\theta/2)$, as discussed above.] 
\begin{figure}
    \includegraphics[width=8cm]{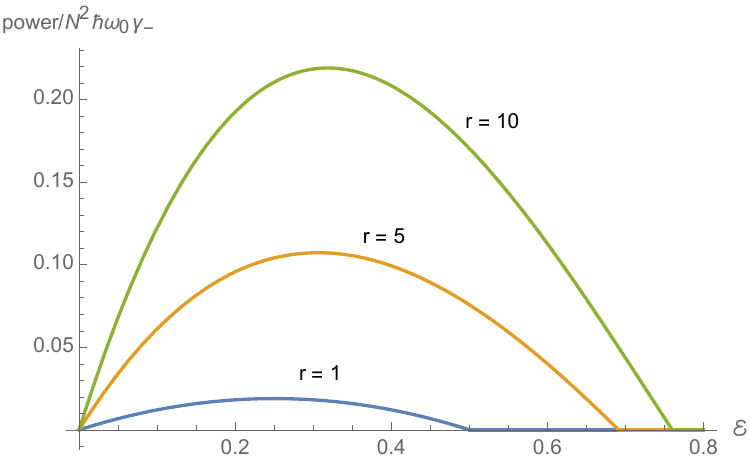}%
    {\caption{Lower bound on average power, divided by $N^2\hbar\omega_0$, vs. maximum final charge, divided by $N\hbar\omega_0$, in the asymptotic limit $N\gg 1$, for $r=1, 5$ and $10$, from the analytical result (\ref{e18}), with $\tau_{90}$ computed numerically (for all curves, we find $\tau_{90} = 2.973 \pm 0.005$). }}
\label{fig:fig3}
\end{figure} 

{\it Energy storage and retrieval.---} It is understood that, in order to store the energy after charging is complete, it would be necessary to somehow suppress the atomic decay channel represented by $\gamma_0$, which implies some further environment manipulation, or some form of ``dynamic coupling'' \cite{dynamic}. Conversely, a simple way to extract the energy would be to reopen that channel, perhaps with the atoms placed within a waveguide to direct the energy.  This collective spontaneous decay would be governed by the master equation (\ref{n8}), with $\gamma=\gamma_0$, and would once again exhibit the superradiant $N$-enhancement of the radiation rate seen in Eq.~(\ref{e15}). Interestingly, and as already noted in \cite{monsel,monsel2} for the single-atom case, the charging procedure considered here allows one to store not only energy, but also coherence.  This can be quantified by the average value of the collective dipole moment $J_+$ in the state $\rho^{(ss)}$ (see Eq.~(\ref{e2})):
\begin{equation}
\av{J_+(0)} = Tr\left[\rho^{(ss)}(\cos\theta J_x^\prime -\sin\theta J_z^\prime+iJ_y^\prime)\right] \to\frac{N\hbar}{2}\sin\theta
\end{equation}
for large $N$. The emitted coherent power, in turn, goes as (cf. \cite{monsel})
\begin{equation}
\frac{dW}{dt} = \frac{\gamma_0}{\hbar} \omega_0 \left|\av{J_+(t)}\right|^2
\label{e19}
\end{equation}
Numerical integration of (\ref{n8}) and (\ref{e19}) (see Fig.~4) indicates that, even for small $N$, a large fraction of the stored energy could be extracted as coherent energy (that is, as the field radiated, coherently, by a ``macroscopic'' dipole) in this straightforward way, again over a short time scaling as $1/N\gamma_0$.  In the large $N$ limit (the extensive regime), solving the equations analytically with the same approximations as before shows that essentially all the stored energy is in principle extractable this way. We intend to study other possible energy extraction methods, similar to those proposed in \cite{monsel}, in future work.  (Note also, in this context, the recent study \cite{asaoka} of a collection of excited two-level atoms as a means to amplify an incident coherent-state field.)

\begin{figure}
    \includegraphics[width=8.5cm]{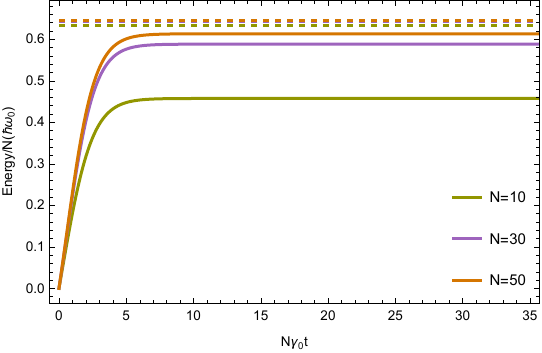}%
    {\caption{Dashed lines: energy stored in steady state for $r =10$, $\theta=1.87$ (from exact result (9)).  Solid lines: Energy emitted coherently by spontaneous emission, from numerical integration of (\ref{e19}).  }}
\label{fig:fig4}
\end{figure}

{\it The dissipative advantage in the extensive regime.---}
As we have seen, in what we have called the extensive regime, our system exhibits a number of properties associated with the thermodynamic limit: the energy is extensive, fluctuations $\Delta E/\av E$ become negligible, and the ergotropy and the total energy coincide (so all the energy is extractable as ``work''; we assume zero temperature, so all the energy should be free energy).  In this regime, the collective advantage could hold over a wide range of values of $N$, possibly even over several orders of magnitude.  Nevertheless, it is important to realize that \emph{no} collective advantage, for any physically realizable system, could survive in a ``thermodynamic limit'' defined as $N\to\infty$, $V\to\infty$, $N/V=$ constant: ultimately, the performance of a battery cannot be affected by the presence of any other batteries outside its light cone, and this is bounded, in the time dimension, by the charging time \cite{gogyan}. We believe, therefore, that the extensive regime (with $V$ and $N$ finite, but $N$ large), is more meaningful than the formal ``thermodynamic limit''---not just for Dicke batteries, but for any kind of quantum battery.

Under the extensivity assumption,  the authors of \cite{campaioli2} used the quantum speed limit (QSL) to derive two possible constraints on the charging power scaling of quantum batteries.  Our dissipative system obeys the second of these ($N$ scaling, derived from the $T > \hbar/{\av E}$ form of the QSL) but appears to violate the first one ($\sqrt N$ scaling, derived from $T > \hbar/{\Delta E}$).  However, this particular form of the QSL (known as the Mandelstam-Tamm, or MT, form) depends in an essential way on the assumption that the evolution is Hamiltonian, and can take a very different form for a dissipative process.  In particular, if we look at just the decay process (\ref{n8}) and (\ref{e19}), and use the result in \cite{huelga}, we obtain a lower bound on the evolution time that scales as $1/\gamma N$.  We expect essentially the same will be the case for atoms confined to a waveguide or a a bad cavity.  On the other hand, for atoms in a good cavity, in the limit where the coupling to the cavity field makes the coupling to other (vacuum) modes negligible, we expect the MT constraint to approach the Hamiltonian result $T\sim 1/\sqrt N$.  As we mentioned in the Introduction, this is, in fact, the scaling exhibited by all the Hamiltonian  proposals for Dicke batteries, and even by the dissipative system demonstrated in \cite{quach} in the extensive regime, which, for that system, was precisely obtained when the collective cavity-mediated light-matter coupling dominated over the decay channels.

Similarly, the constraint $P \le 2\sqrt{\Delta H_B^2\Delta H_C^2}$, derived for Hamiltonian evolution in \cite{rossini} (where $H_B$ and $H_C$ are the  battery and charging Hamiltonians, respectively), does not apply to our system, but the stronger inequality $P(t) \le \sqrt{\Delta H_B^2(t) I_E(t)}$ involving the Fisher information, $I_E$, of the battery alone, derived in \cite{julia}, does hold.  Again for the simpler-to-calculate discharge process, we have found numerically that $\Delta H_B^2$ scales linearly with $N$, whereas $I_E(t)$, a measure of the speed of the evolution of the system in energy space, scales as $N^3$.

In conclusion, we have shown that the state preparation method of Monsel et al., extended to $N$ atoms, maximizes the charging (and discharging) power scaling for a dissipative quantum battery of $N$ non-interacting two-level atoms, in the extensive regime, and exceeds that of all Hamiltonian protocols for the same system.  Our work thus establishes, in principle, a clear advantage of dissipative schemes over Hamiltonian ones.  

On the other hand, something that may need to be kept in mind for practical applications is that all dissipative schemes are necessarily wasteful.  In our case, we note that  during the charging process photons must be scattered, at the frequencies $\omega_0\pm\Omega_P$, at a rate $\sim N^2\gamma_\pm$, over a time of order several times $1/N \Gamma$; hence, the procedure ``wastes'' at least several times as much energy as it stores.  One can say, in this sense, that this charging scheme trades efficiency for power---a tradeoff also familiar from classical thermodynamics.

We acknowledge the MonArk NSF Quantum Foundry supported by the National Science Foundation Q- AMASE-i program under NSF award No. DMR-1906383.

\pagebreak
\widetext
\begin{center}
\textbf{\large Supplemental Materials: calculation of $\av{E(t)}$ for large $N$}
\end{center}
\setcounter{equation}{0}
\setcounter{figure}{0}
\setcounter{table}{0}
\setcounter{page}{1}
\makeatletter
\renewcommand{\theequation}{S\arabic{equation}}
\renewcommand{\thefigure}{S\arabic{figure}}
\renewcommand{\bibnumfmt}[1]{[S#1]}
\renewcommand{\citenumfont}[1]{S#1}

\noindent To calculate the average energy at any time, we note that $J_z$ can be written in terms of the rotated operators as $e^{i J_y\theta/\hbar} J_z^\prime e^{-i J_y\theta/\hbar} = \cos\theta J_z^\prime + \sin\theta J_x^\prime$, and therefore
\begin{equation}
\av{H_0} = \omega_0 \cos\theta\, \av{ J_z^\prime}+\omega_0\sin\theta \,\av{J_x^\prime} 
\end{equation}
where the expectation values can be calculated from Eq.~(7), which is written in terms of the rotated operators (with $H_1$ given by Eq.~(3)).  In the $\{\ket{e_m}\}$ basis, the first term on the right-hand side of (S1) involves only diagonal matrix elements, is independent of $\Omega_P$, and yields the steady-state solution, whereas the second term involves one-off-diagonal elements and gives the transient oscillations at frequency $\Omega_P$.  Specifically, we have, for the first term,
\begin{align}
\frac{1}{\hbar} \frac{d}{dt}\av{ J_z^\prime} &= \sum_{n=-N/2}^{N/2} n \frac{d}{dt} \rho_{nn} \equiv \frac{d}{dt}\av n \cr
&=\sin^4 \frac \theta 2 \gamma_-\sum_{n=-N/2}^{N/2} n\left[ \left(\frac N 2 \left(\frac N 2 +1\right) +n(n-1)\right)\rho_{n-1,n-1} - \left(\frac N 2 \left(\frac N 2 +1\right) +n(n+1)\right)\rho_{nn} \right]\cr
&\quad+\cos^4 \frac \theta 2 \gamma_+\sum_{n=-N/2}^{N/2} n\left[ \left(\frac N 2 \left(\frac N 2 +1\right) +n(n+1)\right)\rho_{n+1,n+1} - \left(\frac N 2 \left(\frac N 2 +1\right) +n(n-1)\right)\rho_{nn} \right] 
\end{align}
where we have defined   $\av{n} = \sum_{n=-N/2}^{N/2} n\rho_{nn}$.  Further defining $\av{n^2} = \sum_{n=-N/2}^{N/2} n^2\rho_{nn}$, and collecting terms, this simplifies to
\begin{equation}
 \frac{d\av n}{dt} = \sin^4 \frac \theta 2 \gamma_-\left[(x-1)\frac N 2 \left(\frac N 2 +1\right)  +(x-1)\av{n^2} -(x+1)\av n\right]
\end{equation}
with $x\equiv r\cot^4(\theta/2)$, and $r=\gamma_+/\gamma_-$, as in the main text.  This can be solved with the approximation $\av{n^2} \simeq \av n^2$, which works in the asymptotic limit $N\gg 1$ because both $\av n^2$ and $\av{n^2}$ are of the order of $N^2$, whereas the difference $\av{n^2} -\av n^2$ is only (at most) of the order of $N$.  The solution is $\av{n(t)} = a - b \tanh\left(c t + \phi_0\right)$, with 
\begin{align}
\av{n(t)} &= a - b \tanh\left(c t + \phi_0\right) \cr
\text{with} &\quad a  = \frac 1 2 \, \frac{x+1}{x-1} \cr
&\quad b = \sqrt{\frac N 2 \left(\frac N 2 +1\right) + a^2} \simeq \frac N 2 \cr
&\quad c =\left( \gamma_+\cos^4 \frac \theta 2 - \gamma_-\sin^4 \frac \theta 2 \right ) b \simeq  \frac N 2  \left( \gamma_+\cos^4 \frac \theta 2 - \gamma_-\sin^4 \frac \theta 2 \right ) \cr
&\quad \phi_0 = \tanh^{-1}[(a - \av{n}_0)/b] = \tanh^{-1}[(a + (N/2) \cos\theta)/b]  \simeq \tanh^{-1} (\cos\theta)
\label{n23}
\end{align}
where the approximations indicated hold in the $N\gg 1$ limit.  Here $\av n_0$ stands for the value of $\av{n}$ at $t=0$; note that this has to be evaluated in the rotated basis, i.e., assuming the initial state is the eigenstate $\ket{-N/2}$ of $J_z$, we have
\begin{equation}
\av n_0 = \frac 1\hbar \av{J_z^\prime}_0 = \bra{-N/2} (\cos\theta J_z -\sin\theta J_x)\ket{-N/2} = -\frac N 2 \cos\theta
\end{equation}
With these approximations, and neglecting $a$ versus $b$ in (S4), we can write, for the first term in (S1),
\begin{equation}
\omega_0\cos\theta \av{J_z^\prime} \simeq  -\frac{N \hbar\omega_0}{2} \tanh\left[\frac N 2\Gamma t + \phi_0 \right]
\end{equation}
with $\Gamma$ defined as in Eq.~(15) of the main text.

As for the second term in (S1), we have
\begin{equation}
\frac 1 \hbar \av{J_x^\prime} =  \sum_{n=-N/2}^{N/2} \left( \frac 1 2 \sqrt{(j(j+1)-n(n-1)} \rho_{n-1,n} + \frac 1 2 \sqrt{(j(j+1)-n(n+1)} \rho_{n+1,n} \right) 
\end{equation}
We can define the sum
\begin{equation}
y \equiv  \sum_{n=-N/2}^{N/2} \sqrt{(j(j+1)-n(n-1)} \rho_{n-1,n}
\end{equation}
and observe that, by relabeling the summation indices, the right-hand side of (S7) reduces to  $Re(y)$.  We then find, in the same way as before, the differential equation for $y$:
\begin{align}
\frac{dy}{dt} 
&=i \sqrt{\Delta^2 + \Omega_R^2}\,y -\frac 1 8 \sin^2\theta\, \gamma_0 y-\sin^4 \frac \theta 2 \gamma_-\sum_{n=-N/2}^{N/2} n \sqrt{(j(j+1)-n(n-1)} \rho_{n-1,n} \cr
&\quad+\cos^4 \frac \theta 2 \gamma_+\sum_{n=-N/2}^{N/2} (n-1)  \sqrt{(j(j+1)-n(n-1)} \rho_{n-1,n}
\label{n25}
\end{align}
To convert this into a closed equation for $y$, we can replace the summation variable $n$ in the last two terms by its average, $\av n$, as calculated above.  We then have 
\begin{equation}
\frac{dy}{dt} =i \sqrt{\Delta^2 + \Omega_R^2}\,y -\frac 1 8 \sin^2\theta\, \gamma_0 y-\left(\sin^4 \frac \theta 2 \gamma_-\cos^4 \frac \theta 2 \gamma_+\right) \bigl(\av n -1 \bigr) y 
\label{n26}
\end{equation}
with $\av{n(t)}$ given by Eqs.~(\ref{n23}).  Note the factor multiplying the $\av n$ term in (\ref{n26}) is the same as $c/b$ in Eqs.~(\ref{n23}), which simplifies somewhat the integration of (\ref{n26}).  The final result is
\begin{equation}
y(t) = y_0 \exp\left[ i \sqrt{\Delta^2 + \Omega_R^2}\,t -\frac 1 8 \sin^2\theta\, \gamma_0 t + \frac{c}{2b} \left(\frac{3-x}{x-1}\right) t \right] \frac{\cosh \phi_0}{\cosh\left(c t + \phi_0\right)}
\label{n27}
\end{equation}
The initial value $y_0$ can be easily obtained from Eq.~(S1), since we must have, for the initial energy,
\begin{equation}
\frac{\av{H_0}_0}{\hbar\omega_0} = -\frac N 2 = \cos\theta \, \av n_0 + \sin\theta \,y_0 = -\frac N 2 \cos^2\theta +\sin\theta \, y_0
\end{equation}
which means $y_0 = -(N/2)\sin\theta$.

Equation (\ref{n27}) can still be simplified somewhat in the large $N$ limit, since it represents a term that goes to zero for $t \sim 1/c \sim 1/N\Gamma$.  This means the second and third term in square brackets become negligible for sufficiently large $N$.  The result is 
\begin{equation}
\frac 1 \hbar \av{J_x^\prime} = \text{Re}(y(t)) \simeq -\frac N 2 \sin\theta\cos(\Omega_P t) \frac{\cosh \phi_0}{\cosh\left(N\Gamma t /2+ \phi_0\right)}
\end{equation}
Our final result is therefore
\begin{equation}
\frac{1}{N\hbar \omega_0} \av{H_0} = -\frac{1}{2} \cos\theta\tanh\left[\frac N 2\Gamma t + \phi_0 \right] -\frac 1 2 \sin^2\theta\cos(\Omega_P t) \frac{\cosh \phi_0}{\cosh\left(N\Gamma t /2+ \phi_0\right)}
\end{equation}
which can be written in a slightly more compact form noting that $\cosh\phi_0 =  \cosh[\tanh^{-1} (\cos\theta)]=1/\sin\theta$. For the energy in the battery (taking the zero of energy in the ground state, before charging begins), this yields
\begin{equation}
\frac{1}{N\hbar \omega_0} \av{E} = \frac 1 2 -\frac{1}{2} \cos\theta\tanh\left[\frac N 2\Gamma t + \phi_0 \right] -\frac 1 2 \sin\theta\frac{\cos(\Omega_P t) }{\cosh\left(N\Gamma t /2+ \phi_0\right)}
\end{equation}

\end{document}